\documentclass[aps,amsmath,twocolumn,a4paper,jcp]{revtex4}

\usepackage{graphicx}
\usepackage{amssymb}

\newcommand {\bee} {\begin{equation}}
\newcommand {\eee} {\end{equation}}
\newcommand {\bea} {\begin{eqnarray}}
\newcommand {\eea} {\end{eqnarray}}
\newcommand {\bes} {\begin{displaymath}}
\newcommand {\ees} {\end{displaymath}}
\newcommand {\beas} {\begin{eqnarray*}}
\newcommand {\eeas} {\end{eqnarray*}}
\newcommand {\comment}[1]{}
\newcommand {\ctwo} {{HT}}
\newcommand {\cone} {{LT}}
\newcommand {\drho} {\rho}
\newcommand {\prho} {\varrho}

\newcommand{\llangle}{\mathbf{<}}
\newcommand{\rrangle}{\mathbf{>}}

\begin{document}

\title{Dissimilar bouncy walkers}

\author{Michael A. Lomholt}
\affiliation{MEMPHYS - Center for Biomembrane Physics, Department of Physics and Chemistry,
University of Southern Denmark, Campusvej 55, 5230 Odense M, Denmark}

\author{Ludvig Lizana}
\affiliation{Niels Bohr Institute, Blegdamsvej 17, DK-2100, Copenhagen,
Denmark}  

\author{Tobias Ambj\"ornsson$^*$}
\affiliation{Department of Astronomy and Theoretical Physics, Lund University,
  S\"olvegatan 14A, SE-223 62 Lund, Sweden, \& Department of Chemistry,
  Massachusetts Institute of Technology, 77 Massachusetts Avenue, Cambridge,
  MA 02139, USA}
\email{tobias.ambjornsson@thep.lu.se}

%
%

\begin{abstract}

  We consider the dynamics of a one-dimensional system consisting of
  {\em dissimilar} hardcore interacting (bouncy) random walkers. The
  walkers' (diffusing particles') friction constants $\xi_n$, where
  $n$ labels different bouncy walkers, are drawn from a distribution
  $\prho(\xi_n)$. We provide an approximate analytic solution to this recent
  single-file problem by combining harmonization and effective medium
  techniques. Two classes of systems are identified: when
  $\prho(\xi_n)$ is heavy-tailed, $\prho(\xi_n)\simeq
  \xi_n^{-1-\alpha} \ (0<\alpha<1)$ for large $\xi_n$, we identify a
  new universality class in which density relaxations, characterized
  by the dynamic structure factor $S(Q,t)$, follows a Mittag-Leffler
  relaxation, and the the mean square displacement of a tracer
  particle (MSD) grows as $t^\delta$ with time $t$, where
  $\delta=\alpha/(1+\alpha)$. If instead $\prho$ is light-tailed
  such that the mean friction constant exist, $S(Q,t)$ decays
  exponentially and the MSD scales as $t^{1/2}$. We also
  derive tracer particle force response relations. All results are
  corroborated by simulations and explained in a simplified model.
\end{abstract}

\maketitle

%
%

\section{Introduction}

The staggering motion of a lone drunk has, since the classic
correspondence \cite{Pearson,Lord_Rayleigh} between Karl Pearson and
Lord Rayleigh in 1905, been a well established metaphor for a random
walk. This view was embraced in a celebrated work \cite{Fisher_84} by
Michael E. Fisher in which he discussed statistical aspects of many
drunks. In particular, Fisher introduced the notion of harmless drunks
for identical non-interacting walkers, vicious drunks representing the
motion of antagonists, and bouncy walkers exhibiting hardcore
repulsive interactions in one dimension.

The bouncy walker problem, which we here revisit, finds a number of
experimental realizations: transport in microporous
materials \cite{KKDGPRSUK, meersmann00,hahn96} (e.g. zeolites),
colloidal systems \cite{WBL}, molecular sieves \cite{gupta95} and
biological pores \cite{HOKE}. Hard core repulsion of binding proteins
diffusing along DNA was recently suggested to be important in
transcription. \cite{Elf_09}

On the theoretical side, the many-body problem of identical bouncy walkers in
one dimension, also referred to as single-file diffusion \cite{LE} or
symmetric exclusion \cite{Arratia_83}, has received considerable attention
since it was introduced by Harris\cite{HA} in 1965. The main result for
thermal initial conditions is that a tracer particle exhibits sub-diffusion
\cite{LE,HA,Alexander_Pincus_78,Kollmann_03,barkai2009theory}; the tracer mean
square displacement (MSD) is proportional to $t^{1/2}$ even though the
collective behavior is diffusive (the structure factor $S(Q,t)$ decays
exponentially \cite{LIAM2}). In this study, we ask: what are corresponding
expressions for the MSD and $S(Q,t)$ for a system of {\it dissimilar}
bouncy walkers? We show that this question can be answered in closed form,
using harmonization \cite{lizana_10} and effective medium approaches
\cite{landauer52,kirkpatrick71,alexander81}. The analytic predictions are
  corroborated by extensive simulations. Our results extend, in ways
described throughout the study, a very limited number of previous studies
\cite{Jara_Gonzalves, TA_etal,Jara_09,Flomenbom} of bouncy walkers of
different type.

\section{Equations of motion}

We consider strongly overdamped motion of Brownian
particles, in an infinite one dimensional system, interacting via a
two-body short-range repulsive potential. The nearest neighbor
potential ${\cal V}(|x_n(t) - x_{n'}(t)|)$, where $x_n(t)$ is the
position of the $n$th particle, has a hard-core part which excludes
particles from overtaking each other.  The Langevin equations of
motion are thus
\bee\label{eq:LE_first}
\xi_n \dot{x}_n(t) = \sum_{n'}
\mathfrak{f}[x_n(t) - x_{n'}(t)] +\eta_n (t) + f_n (t), 
\eee
where a dot denotes time derivative, $\mathfrak{f}=-\partial {\cal
  V}/ \partial x_n$, $\eta_n(t)$ is a Gaussian zero-mean noise,
$\llangle\eta_n(t)\rrangle =0$, with correlations that are determined
by the fluctuation-dissipation theorem \cite{kubo66} to be
$\llangle\eta_n(t)\eta_{n'}(t')\rrangle
=2k_BT\xi_n\delta(t-t')\delta_{n,n'}$, where $k_B$ is the Boltzmann
constant and $T$ the temperature.  $f_n(t)$ is an external force and
particle $n$ has the friction constant $\xi_n$. In
  Sec. \ref{sec:S_Q_t_main} where the system's collective behavior is
  considered we take $f_n(t)\equiv 0$. At the end of
  Sec. \ref{sec:tracer} and in Sec. \ref{sec:self_averaging} we assume
  $f_n(t)$ to be a static force turned on at $t=0$ or an oscillating force
  acting only on the tracer particle. 

The dissimilarity of the random walkers enter through their different friction
constants, $\xi_n$, which here are assumed to be identically distributed
random variables taken from a probability density $\prho(\xi_n)$.
Eq. (\ref{eq:LE_first}), the noise correlation and $\prho(\xi_n)$ defines the
dissimilar bouncy walkers problem which we deal with by using two
non-equilibrium statistical physics methods outlined below and elaborated in
appendix \ref{sec:EM}.

 We complement all analytic results by stochastic simulations for
  hardcore interacting particles in a box with reflecting boundary
  conditions, detailed in appendix \ref{sec:simulation_details}. The
  box size is chosen sufficiently large such that boundary effects are
  negligible for the center particle within the duration of the simulation run.

A prototypical system where Eq. (\ref{eq:LE_first}) is applicable
  is protein diffusion on DNA molecules\cite{Elf_09}, where the
  heterogeneity in friction constants originate from the different
  proteins varying binding strength to the DNA molecule.

\section{Harmonization and effective medium theory}\label{sec:EMH_main}

\begin{figure}[tb]
\begin{center}
\includegraphics[width=0.7\columnwidth]{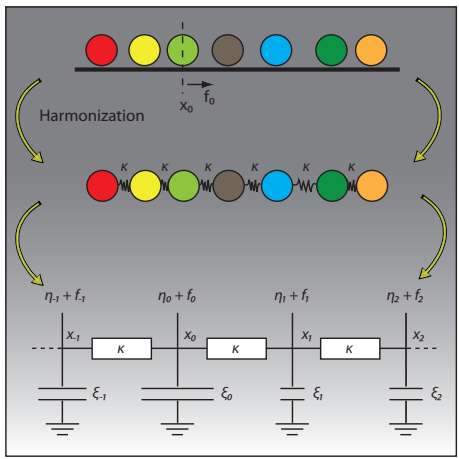}
\end{center}
\caption{Two mappings bring the bouncy walker problem (top) on a tractable
form. Harmonization maps the system onto that of harmonically coupled, 
Rouse chain type \cite{Khokhlov}, beads (middle). The corresponding
Eq. (\ref{eq:khoklov_y}) has the same form as the equation appearing in
resistor-capacitor network theory where $x_n$ represent the node potentials,
$\xi_n$ are capacitances, $\kappa$ is the conductance of the resistors and
$\eta_n+f_n$ is an external current entering the circuit at node $n$
(bottom). This analogy allow us to derive an effective medium
Eq. (\ref{eq:EMH}) of the type originally used to calculate conductances of
binary metallic mixtures \cite{landauer52} and later extended to resistor
networks \cite{kirkpatrick71} and applied to diffusion in random media
\cite{alexander81}.
  \label{fig:mappings}}
\end{figure}
Our first step towards bringing the dissimilar walker problem on a tractable
form is to generalize the {\it harmonization} approach
\cite{lizana_10} to the case when the particles have different friction
constants. In the long time limit this allows us to map the equations of
motion (\ref{eq:LE_first}) onto that for a linear chain of interconnected
springs
\bee
\xi_n\frac{d x_n(t)}{d t}=\kappa\left[x_{n+1}(t)+x_{n-1}(t)-2 x_n(t)\right]+\eta_n(t)+f_n(t)\label{eq:khoklov_y},
\eee
corresponding to harmonically coupled beads, see
Fig. \ref{fig:mappings}.  The effective nearest neighbor spring
constant $\kappa$ is calculated by demanding that the change in free
energy for small particle displacements in the original system equals
the free energy change in the spring system. For hardcore interacting
particles of size $b$ (used in our simulations) this harmonization
procedure yields \cite{lizana_10} $\kappa = \drho^2 k_B T (1 - \drho b
)^{-2}$, where $\drho$ is the particle density.  The harmonization
approach relies on the assumption that local equilibration is faster
than tracer particle dynamics, for sufficiently long times. We find
self-consistently that this holds since, as we will see, the MSD for a
single particle is proportional to $t^{\delta}$ with $\delta\le 1/2$
[see Eq. (\ref{eq:MSD})], i.e., the particle cross a distance of length
$L$ in a time on the order of $L^{2/\delta}$. At long distances this
is indeed slower than the corresponding density relaxation time which
scales as $L^{1/\delta}$ [see Eq. (\ref{eq:S_Q_t})]. 

Second, applying an effective medium approach
\cite{landauer52,kirkpatrick71,alexander81} to the harmonized equations yields
a set of generalized Langevin equations containing a time dependent, but $n$
independent, memory kernel $\xi_{\rm eff} (t)$. This method is based on an
analogy of our system with that of resistor networks, see
Fig. \ref{fig:mappings}. In the continuum limit (in the long-wavelength
  limit $n$ can be treated as a continuum variable\cite{lizana_10}) the
effective medium harmonized equations are
\bee \label{eq:EMH}
\int_{-\infty}^\infty
\xi_{\rm eff}(t-t')
\dot{x}_n(t') \,dt'
=\kappa \frac{\partial^2 x_n(t)}{\partial n^2}
+\eta_n^{\rm eff}(t)+f_n(t).
\eee 
The lower integration limit being $-\infty$ physically
  corresponds to the system dynamics being ``turned on'' in the
  infinite past, thereby effectively bring the system to equilibrium
  at any finite time $t$. As shown below $\xi_{\rm eff}(t-t')=0$ for
  $t'>t$, as required by causality.  The fluctuation-dissipation
theorem enforces a relation between the memory kernel and the
effective noise $\eta_n^{\rm eff}(t)$ \cite{kubo66}:
\bee\label{eq:noise_corr} 
   \left<\eta_n^{\rm
    eff}(t)\eta_{n'}^{\rm eff}(t')\right>=k_B T \xi_{\rm
  eff}(|t-t'|)\delta (n-n'),
\eee 
where the brackets $\langle .. \rangle$ represent an implicit average
over quenched friction constants, besides averages over different
realization of the noise and random initial positions. We label this
average the {\em heterogeneity-averaged} case \cite{Jara_Gonzalves},
which is contrasted by the {\em non-averaged} case (represented by
$\llangle .. \rrangle$) where an average over the probability density
of friction constants is not performed. In the simulations for the
non-averaged case the same $\xi_n$'s are used when averaging over
thermal noise (i.e., for each simulation run). For the heterogeneity
averaged case we draw new friction constants whenever we make a new
initial particle positioning.

The memory kernel, $\xi_{\rm eff}(t)$, in Eq. (\ref{eq:EMH}) is determined by
imposing a self-consistency criterion (see appendix \ref{sec:EM}). We identify
two classes of systems: For light-tailed ({\cone}) systems with a probability
density $\prho(\xi)$ such that the mean friction $\bar{\xi}=\int_0^\infty
\xi\prho(\xi) d\xi$ exists, we obtain, in the long time limit, a memoryless
kernel
\bee
\xi_{\rm eff} (t) \sim {\bar \xi}\delta(t),
\eee
 where $\delta(z)$ is the
Dirac delta-function. Thus, {\cone} systems are in the same universality class
as that of identical bouncy walkers \cite{lizana_10}. In contrast, we find
that heavy-tailed ({\ctwo}) systems, where $\prho(\xi)=A\xi^{-1-\alpha}$ for
large $\xi$, with $0<\alpha<1$ ($A$ is a normalization constant), belong to a
new universality class where the memory kernel has a power-law decay with time
\bee\label{eq:C2_xi_eff}
  \xi_{\rm eff}(t)\sim \chi \frac{t^{-2\delta}}{\Gamma(1-2\delta)}  \theta(t)
\eee
where $\Gamma(z)$ is the gamma function, $\theta(z)$ is the Heaviside
step function, $\chi=(4\kappa)^{1-2\delta}
(A\pi/\sin[\delta\pi/(1-\delta)])^{2(1-\delta)}$ and
  \bee\label{eq:delta}
\delta=\frac{\alpha}{1+\alpha}.
  \eee
  We point out that for HT systems it is only the tail of the distribution
  [the detailed structure of $\prho(\xi)$ for small $\xi$ is not important]
  which determines the long-time dynamics. The equations
  (\ref{eq:EMH})-(\ref{eq:C2_xi_eff}) in this section allow us to calculate
  explicit long-time expressions for observables in dissimilar bouncy walker
  systems [i.e., systems described by Eq. (\ref{eq:LE_first})].

\section{Collective behavior}\label{sec:S_Q_t_main}  

\begin{figure}[tb]
\begin{center}
\includegraphics[width=8.5cm]{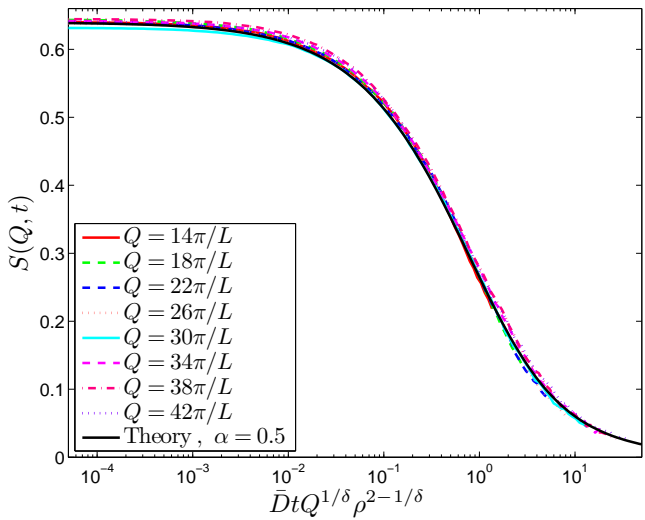}
\end{center}
\caption{Dynamic structure factor. Collapse-plot for $S(Q,t)$ for
  different wavevectors $Q$ as a function of time, illustrating
  anomalous decay.  The black solid curve is the analytic prediction,
  Eq.  (\ref{eq:S_Q_t}). Simulations are non-averaged results from 32000
  noise realizations with $N=501$ particles in a system of length
  L=2501b ($b$ is the particle size) with $\alpha=0.5$. The average
  diffusion constant is denoted $\bar{D}$. 
  Note that, in contrast to tracer particle observables (see inset of
  Fig. \ref{fig:compare}) $S(Q,t)$ is a sum over all particles
  and therefore display self-averaging.  
  \label{fig:SQt}}
\end{figure}

Let us now consider the collective behavior of dissimilar bouncy walker
systems, using the analytic approach in Sec. \ref{sec:EMH_main} and the
simulation scheme from appendix \ref{sec:simulation_details}.  A quantity
capturing the collective motion and easily accessible in experiments
\cite{Berne_Pekora} is the dynamic structure factor $S(Q,t)$. For
translationally invariant systems we use $S(Q,t) = \sum_n \langle
e^{iQ[x_n(t)-x_0(0)]}\rangle$ where the summation index runs over all
particles. For {\cone} systems we find from Eq. (\ref{eq:EMH}) (see appendix
\ref{sec:S_Q_t}), in the limit of small wavevectors $Q\neq 0$ and long times,
that 
\bee
S(Q,t)\sim S(Q,0)\exp(-D_c Q^2 t )
\eee
 with the static structure factor
$S(Q,0)=k_B T\drho^2/\kappa$ and the collective diffusion constant
$D_c=\kappa/(\drho^2\bar{\xi})$. Density relaxations are thus exponential as
for a system of harmless drunks \cite{Berne_Pekora} or identical bouncy
walkers \cite{TA_etal}. For {\ctwo} systems we find a collective behavior
which is drastically different:
\bee\label{eq:S_Q_t} 
 S(Q,t)\sim S(Q,0) E_{2\delta}(-\lambda_c Q^2 t^{2\delta}),\qquad Q\neq 0 
\eee 
where $E_{\alpha}(z)$ is the Mittag-Leffler function \cite{MEKL1} and
$\lambda_c=\kappa/(\drho^2\chi)$ is a generalized collective diffusion
constant. Here, the density relaxations exhibit anomalously slow power-law
decay with time. The anomalous decay of $S(Q,t)$ is illustrated in
Fig. \ref{fig:SQt} together with simulation results for the non-averaged case.

From Eq. (\ref{eq:S_Q_t}) and Onsager's regression hypothesis \cite{onsager31}
it follows that perturbations of the concentration $c(X,t)=\langle \sum_m
\delta(X-x_m(t)) \rangle $ around its equilibrium value $\drho$ decay
according to the fractional diffusion equation $\partial c(X,t)/ \partial
t=\lambda_c D^{1-2\delta}_t\partial^2 c(X,t)/\partial X^2$ where
${}_0D^{1-2\delta}_t$ is the fractional Riemann-Liouville operator. This
equation describes also the  subdiffusive motion of non-interacting
continuous time random walkers (CTRW) with a power-law waiting time density
\cite{MEKL1}. However, the nature of the process is here very
  different. In particular, our system has a stationary fluctuating
  equilibrium state, since the underlying many-body dynamics is Markovian with
  a well-defined equilibrium (two point correlation functions only depend on
  the time difference, i.e., the system do not age). In contrast, CTRWs do not
  have a stationary state (their two point correlation functions age).
Therefore, to the best of our knowledge, dissimilar bouncy walkers are the
first example of a stationary physical system obeying (to a good
approximation) the fractional diffusion equation for density relaxations. The
fact that our {\ctwo} bouncy walkers belong to a different universality class
than CTRW systems is also captured by the difference in tracer particle
behavior.

\section{Tracer particle dynamics, heterogeneity-averaged case}\label{sec:tracer}

\begin{figure}[tb]
\begin{center}
\includegraphics[width=8.5cm]{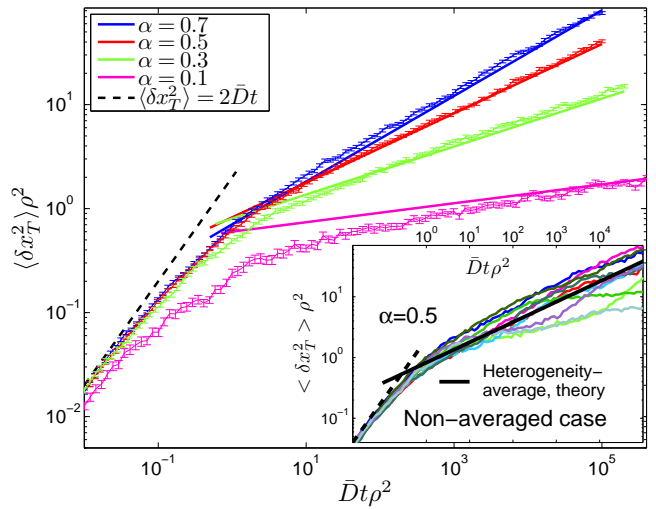}
\end{center}
\caption{Tracer particle mean square displacement. MSD as a function
of time for the heterogeneity-averaged and non-averaged (inset)
cases. The straight solid lines are the asymptotic analytic
prediction, Eq. (\ref{eq:MSD}), and the dashed lines are the result
for an independent random walker. In the heterogeneity-averaged case
 the simulations are averages over 2400 realizations, and the center
particle is taken as tracer particle. We used a box size $L=10001b$
($b$ is the particle size) and $N=1001$ particles for $\alpha=0.7$ and
$\alpha=0.5$. For $\alpha=0.3$ ($\alpha=0.1$) we had $N=501$ and
$L=5001b$ ($N=251$ and $L=2501b$).  The average diffusion constant is
$\bar{D}$.  Inset: Non-averaged MSD
simulations for $\alpha=0.5$ and 800 noise realizations for each
curve.  Notice the spread around the analytic result. 
\label{fig:compare}}
\end{figure}

We now address tracer particle dynamics in the absence of an external
force (see appendix \ref{sec:tracer_MSD} for details) using
Eqs. (\ref{eq:EMH})-(\ref{eq:C2_xi_eff}) . For {\cone} systems we find the
MSD 
\bee
\langle\delta x_\mathcal{T}^2(t)\rangle \sim k_B T \sqrt{\frac{4t}{\pi
  \kappa \bar{\xi}} }\propto t^{1/2},
\eee
for particle $\mathcal{T}(=0)$. Thus, the MSD for {\cone} systems
  take the same form as for identical walkers\cite{lizana_10}, but with the
  friction constant for the homogeneous case replaced by the mean friction
  constant $\bar{\xi}$. The same $t^{1/2}$-scaling was recently obtained for
dissimilar walkers on a lattice. \cite{Jara_Gonzalves,TA_etal} A very
different MSD exponent is found for {\ctwo} systems where instead
\bee\label{eq:MSD} \langle\delta
x_\mathcal{T}^2(t)\rangle \sim \frac{k_B T}{\sqrt{\kappa \chi}}
\frac{t^{\delta}}{\Gamma(1+\delta)}\propto t^\delta
 \eee 
 with $\delta <1/2$ [see Eq. (\ref{eq:delta})] indicating ultra-slow dynamics
 for the tracer particle. Heterogeneity-averaged simulations show excellent
 agreement with this result, see Fig. \ref{fig:compare}. The MSD exponent
 $\delta=\alpha/(1+\alpha)$ has also recently been derived for lattice systems
 in a rather technical mathematical study by Jara \cite{Jara_09} and obtained
 through scaling arguments in Ref. \onlinecite{Flomenbom}. Besides providing a
 simplified derivation for continuum systems, we extend Jara's result by also
 obtaining an explicit expression for the MSD prefactor.  The probability
   density function (PDF) for the tracer particle position is a Gaussian
   (since the noise is taken from a multi-variate normal distribution) with a
   width described by Eq. (\ref{eq:MSD}), see appendix
   \ref{sec:simulation_details} for corroborating simulations.

\begin{figure}[tb]
\begin{center}
\includegraphics[width=8.5cm]{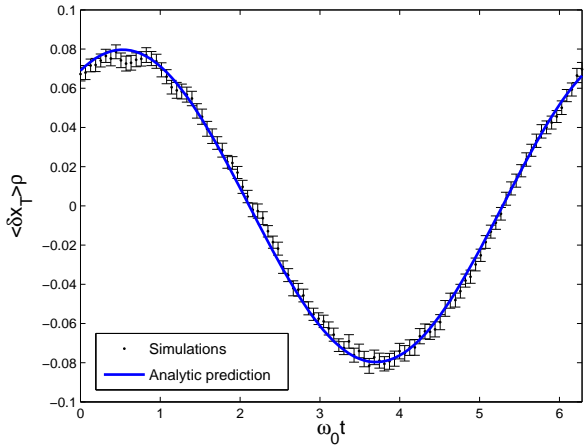}
\end{center}
\caption{Mean
    displacement for a tracer particle in the presence of a time-varying
    force. Simulations for the average displacement of a tracer particle as
  a function of time within a single period of a harmonically oscillating
  external force, heterogeneity-averaged case.  Parameters are $\alpha=0.5$, $N=501$,
  $L=5001b$ ($b$ is the particle size), $\alpha_0=0.002$ and $\omega_0/(2\pi)=
  10^{-5}$. The average is over last 100 periods of the force for 1280
  noise realizations each lasting 200 periods. 
  The blue solid curve is the analytic
  prediction, Eq. (\ref{eq:force_resp}). Notice the excellent agreement
  between theory and simulations.
  \label{fig:force}}
\end{figure}

In the presence of a external force on the tracer particle we calculate
force-response relations (see Appendices \ref{sec:force} and
\ref{sec:gen_Einstein}). For the case of a static force $f_n(t) =
\delta_{n,0}F_0\theta(t)$ with magnitude $F_0$, we show by explicit
calculation that the generalized Einstein relation $ \langle
x_\mathcal{T}(t)\rangle_f= \langle\delta x_\mathcal{T}^2(t)\rangle F_0/(k_B
T)$ is satisfied for both {\cone} and {\ctwo} systems. The bracket $\langle
..\rangle_f$ denotes an average in the presence of $f_n(t)$. For an
oscillatory force $f_n(t)=\delta_{n,0}F_0\cos(\omega_0 t)$ with angular
velocity $\omega_0$ we get $\langle
x_\mathcal{T}(t)\rangle_f=[F_0/(2\sqrt{\kappa \bar{\xi} \omega_0})]
\cos(\omega_0 t-\phi)$ where $\phi=\pi/4$ for {\cone} systems. For {\ctwo}
systems we find
\bee \langle
x_\mathcal{T}(t)\rangle_f=\frac{F_0}{2\sqrt{\kappa \chi}
  \omega_0^{\delta}}\cos(\omega_0 t-\phi)
\label{eq:force_resp}
\eee
with the non-trivial phase-shift $\phi=\delta \pi/2<\pi/4$, which
compactly quantifies the subtle interplay between bounciness and the
degree of friction dissimilarity. Eq. (\ref{eq:force_resp}) is
compared to simulations for the heterogeneity-averaged case in Fig. \ref{fig:force}.

\section{Self-averaging}\label{sec:self_averaging}

Let us comment on the difference between the non-averaged and
heterogeneity-averaged cases. From simulations for both cases we found
that $S(Q,t)$ and the MSD are self-averaging quantities for the
{\cone} class, as has been previously shown for lattice systems
\cite{Jara_Gonzalves,TA_etal}. For {\ctwo} systems $S(Q,t)$
self-averages while the MSD does not (see Figs. \ref{fig:SQt} and
\ref{fig:compare}). In order to understand this in more detail, we put
forward a simplified model for tracer particle dynamics in the
non-averaged case: under the influence of a constant force, $F_0$, we
write $\xi_{\rm tot} d\llangle \delta
x_\mathcal{T}\rrangle_{F_0}/dt=F_0$.  We use brackets
$\llangle..\rrangle$ to denote averages over thermal noise and random
initial positions for the non-averaged case. The quantity $\xi_{\rm
  tot}$ represents the total friction experienced by the particle
which is not only its own, but also that of an approximate number $N$
of its neighbors on which it is ``pushing'', i.e. $\xi_{\rm
  tot}=\xi_0+\xi_1+\xi_2+...+\xi_N$. Within the harmonization approach
\cite{lizana_10} we have approximately that  the effective spring
  constant for $N$ serially connected springs is $\kappa/N$, i.e., we
have $F_0=(\kappa/N) \llangle \delta x_\mathcal{T}\rrangle_{F_0}$,
from which $N=\kappa \llangle \delta x_\mathcal{T}\rrangle_{F_0}/F_0
$; the larger the displacement the larger is the number of particles
contributing to $\xi_{\rm tot}$.  Using this result and multiplying
and dividing the equation of motion above by $N^{1/\alpha}$ we arrive
at
\bes \underbrace{\frac{\xi_1+\xi_2+...+\xi_N}{N^{1/\alpha}}}_{\zeta} (\kappa
\llangle \delta x_\mathcal{T}\rrangle_{F_0}/F_0 )^{1/\alpha} \frac{d\llangle
  \delta x_\mathcal{T}\rrangle_{F_0} }{dt} =F_0, 
\ees
where $\zeta$ is a random variable with an $N$-independent distribution
$\psi(\zeta)$ for $N\rightarrow \infty$. Solving the equation above and
employing the fluctuation-dissipation theorem in the form of the generalized
Einstein relation, $k_B T \llangle \delta
x_\mathcal{T}\rrangle_{F_0}/F_0=\llangle \delta x_\mathcal{T}^2\rrangle$, we
find the MSD for the non-averaged case:
\bee\label{eq:msd_quenched} \llangle
\delta x_\mathcal{T}^2\rrangle\propto \frac{k_B T}{\zeta^\delta
  \kappa^{1-\delta}}t^\delta.
\eee 
In agreement with the non-averaged case simulations (inset Fig.
\ref{fig:compare}), the MSD prefactor is a random variable $\zeta$. Thus,
{\ctwo} systems do not show self-averaging, rather each new realization of
$\xi_n$'s gives a different fluctuating prefactor for the MSD (but with the
same scaling with time).
 The very simplistic argument here predicts that
 for {\ctwo} systems $\psi(\zeta)$ is a one-sided L\'evy stable
distribution \cite{MEKL1}, but further work is needed to pinpoint the
correct functional form for $\psi(\zeta)$.
 We point out that the non-averaged {\ctwo}  case has not been
investigated in previous studies \cite{Jara_09}. Using the fact that
$\langle \zeta^{-\delta}\rangle \propto A^{-(1-\delta)}$ for the
power-law distribution of friction constants considered here, our
simplified model recovers Eq. (\ref{eq:MSD}), up to a dimensionless
$\alpha$-dependent prefactor for the heterogeneity-averaged case. For {\cone}
systems ($\alpha=1$) the central limit theorem gives
$\psi(\zeta)=\delta(\zeta-\bar{\xi})$ as $N\rightarrow \infty$, i.e.,
the argument above predicts that {\cone} systems possess the
self-averaging property as indeed found in our
simulations.\cite{TA_etal}

\section{Summary and outlook.}

We investigated the dynamics of a one-dimensional system consisting of
{\em dissimilar} bouncy random walkers, with friction constants drawn
from a probability density. Two classes of systems were identified:
those with heavy-tailed ({\ctwo}) probability densities in which density
relaxations, characterized by the dynamic structure factor $S(Q,t)$,
follows a Mittag-Leffler relaxation. The mean square displacement of a
tracer particle (MSD) grows as $t^\delta$ with time $t$, where
$\delta<1/2$, and we find a phase-shift $\phi=\delta\pi/2$ in the
force-response relation for such systems. For light-tailed ({\cone})
probability densities, $S(Q,t)$ decay exponentially, the MSD scales as $t^{1/2}$
and a phase-shift $\phi=\pi/4$ in the force response is obtained. We
also introduced a simplified model which allowed us to address the
problem of self-averaging for {\cone} and {\ctwo} systems. All results were
corroborated by extensive simulations.

Prior to this study, dissimilar hardcore interacting diffusing particles have
been investigated only in a few publications, see Refs.
\onlinecite{Jara_Gonzalves,TA_etal,Flomenbom,Jara_09}. Using
harmonization and effective medium approaches we here extended previous
results by providing: (i) an explicit expression for the structure factor
$S(Q,t)$; (ii) an explicit prefactor for the MSD for {\ctwo} systems; (iii)
force-response relations for a tracer particle; (iv) insights concerning the
difference between non-averaged and heterogeneity-averaged cases for {\ctwo}
systems; (v) extensive simulation results. We further provided a much
simplified framework, Eqs. (\ref{eq:EMH})-(\ref{eq:C2_xi_eff}), within which
further observables, such as correlation functions between particles (see
Ref. \onlinecite{lizana_10}), can be calculated.

Solvable many-body problems have served as important models for real systems
over the last century. The problem of dissimilar bouncy walkers
is expected to find applications in systems
such as passive diffusion of proteins along DNA or particles diffusing in 
nanochannels. Further, often the dynamics of macromolecules in living cells is
found to be subdiffusive \cite{weiss_09}. The
techniques developed in this study may be of use also in such higher
dimensional heterogeneous systems.

%
%

\section{Acknowledgment} 

We are grateful to Mehran Kardar and Eli Barkai for discussions and Milton
Jara for helpful correspondence regarding Ref. \onlinecite{Jara_09}. T.A. and
L.L. acknowledges funding from the Knut \& Alice Wallenberg Foundation. T.A. is
grateful for funding from the Swedish Research Council. Computer time was
provided by the Danish Center for Scientific Computing.

\appendix

\section{Effective medium approach.}\label{sec:EM}

Our second step in deriving a
set of manageable equations for the dissimilar bouncy walker problem is to
introduce an effective medium approximation known for instance from the theory
of resistor networks \cite{kirkpatrick73}, see Fig. \ref{fig:mappings}. In the
language of single-file motion the approximation consists of replacing the
different frictions $\xi_n$ in Eq. (\ref{eq:khoklov_y}) with an effective
friction, or memory kernel, $\xi_{\rm eff}(t)$, which is identical for all the
particles, but instead has a memory. The memory kernel $\xi_{\rm eff}(t)$ is
chosen such that if one takes a random particle from the harmonized system
with friction constant $\xi_n$ and place it among particles characterized by
$\xi_{\rm eff}(t)$, then the mobility of this particle is, on average
[averaged using $\prho(\xi_n)$], required to be identical to the mobility of a
tracer particle in the system where all particles had friction $\xi_{\rm
  eff}(t)$. A detailed calculation of $\xi_{\rm eff}(t)$ is given below.

Let us, as our starting point, consider the harmonized equation
(\ref{eq:khoklov_y}), which we write:
\bee
\xi_n\frac{d y_n(t)}{d t}=\kappa\left[y_{n+1}(t)+y_{n-1}(t)-2 y_n(t)\right]+\eta_n(t)+f_n(t)\label{eq:khoklov_y2}
\eee
where we introduced $y_n(t)=x_n(t)-n\drho^{-1}$, which eliminates the average
values so that $\llangle y_n(t)\rrangle=0$.  Our goal here is to find the
friction kernel $\xi_{\rm eff}(t)$ entering the effective medium harmonization
equation of motion
\bea \label{eq:eff_medium_eq_of_motion}
\int_{-\infty}^\infty\hspace{-0.15cm} \xi_{\rm eff}(t-t')\frac{d y_n(t')}{d
t'}dt'&\hspace{-0.15cm}=&\hspace{-0.15cm}\kappa\left[y_{n+1}(t)+y_{n-1}(t)-2 y_n(t)\right]\nonumber\\
&&+\eta_n^{\rm eff}(t)+f_n(t)
\eea 
which in the continuum limit with respect to $n$ becomes Eq.
(\ref{eq:EMH}) in the main text.  We introduce the frequency-dependent
generalized friction constants
\bee
\gamma^{(\pm)}_n(\omega)\equiv -\frac{F_{n\pm 1\to n}(\omega)}{\llangle v_n(\omega)\rrangle}\label{eq:gammadef}
\eee
where $v_n(\omega)=-i\omega y_n(\omega)$ is the Fourier transform of
the velocity, and $F_{n\to n+1}=-\kappa(y_{n+1}-y_n)$ is the force
from particle $n$ on particle $n+1$ within the harmonization
approach. The quantity $\gamma^{(+)}_n$ ($\gamma^{(-)}_n$) can be
thought of as a generalized friction of particle $n$ due to the part
of the harmonic chain lying to right (left) of particle $n$. We here
define the Fourier-transform with respect to time of a function $A(t)$
according to: $A(\omega)=\int_{-\infty}^\infty e^{i\omega t}A(t)dt$,
with inverse transform $A(t)=\int_{-\infty}^\infty e^{-i\omega t}
A(\omega) d\omega/(2\pi)$. Using the generalized frictions introduced
above and setting $f_n(t)=f_0(t)\delta_{n,0}$ one readily obtains,
from the Fourier-transform of Eq. (\ref{eq:khoklov_y}), the mobility
 of particle 0
\bee\label{eq:mu_0}
\mu_0(\omega)\equiv \frac{\langle v_0(\omega)\rangle}{f_0(\omega)}=\frac{1}{\xi_0+\gamma^{(+)}_0(\omega)+\gamma^{(-)}_0(\omega)},
\eee
as well as the recursion relations
\bee\label{eq:recursion1}
\gamma^{(+)}_n(\omega)=\frac{\xi_{n+1}+\gamma^{(+)}_{n+1}(\omega)}{1-i\omega \left[\xi_{n+1}+\gamma^{(+)}_{n+1}(\omega)\right]/\kappa}\; ,\quad n\ge 0
\eee
and
\bee\label{eq:recursion2}
\gamma^{(-)}_n(\omega)=\frac{\xi_{n-1}+\gamma^{(-)}_{n-1}(\omega)}{1-i\omega
  \left[\xi_{n-1}+\gamma^{(-)}_{n-1}(\omega)\right]/\kappa}\; ,\quad n\le
0
\eee
So far everything is only a reformulation of Eq. (\ref{eq:khoklov_y2}) for the
case $f_n(t)=f_0(t)\delta_{n,0}$. We point out that we could equally well have
reformulated Eq. (\ref{eq:khoklov_y2}) by using Laplace-transformed quantities
$A(s)=\int_0^\infty e^{-st}A(t)dt$; the corresponding recursion relations
would then take the same form as in Eqs. (\ref{eq:recursion1}) and
(\ref{eq:recursion2}), with $s=-i\omega$. However, for the purpose of avoiding
to explicitly compute averages over initial positions using Fourier-transforms
is more convenient, and we therefore employ Fourier-transform techniques
throughout this study.

Let us now turn to the problem of computing $\xi_{\rm eff}(t)$ in the
effective medium Eq. (\ref{eq:eff_medium_eq_of_motion}). We invoke the
following procedure: (A) Fourier-transform Eq. (\ref{eq:khoklov_y2}), and
replace all friction constants for $n\neq 0$ by an $n$ independent but
frequency dependent effective friction, i.e. we let
\bee\label{eq:xi_n} \xi_n\rightarrow (1-\delta_{n,0})\xi_{\rm
  eff}(\omega)+\xi_0\delta_{n,0}.  
\eee 
From these equations the mobility $\mu_0(\xi_0)$ for particle $0$ is
determined.  (B) We then obtain $\xi_{\rm eff}(\omega)$ by imposing the
self-consistency requirement that $\mu_0(\xi_0)$ averaged over different
realizations of $\xi_0$,
\bee\label{eq:EM_criterion} \bar{\mu}_0=\int_0^\infty \mu_0(\xi_0) \prho(\xi_0)
d\xi_0, 
\eee 
is equal to the tracer particle mobility, $\mu_{\rm eff}(\omega)$, obtained
from the effective medium Eq. (\ref{eq:eff_medium_eq_of_motion}), i.e.,
when also $\xi_0$ is replaced by $\xi_{\rm eff}(\omega)$.

Let us now carry out the scheme above. (A) From Eqs. (\ref{eq:recursion1}),
(\ref{eq:recursion2}) and (\ref{eq:xi_n})
\bee
\gamma_{\rm eff}=\frac{\xi_{\rm eff}+\gamma_{\rm eff}}{1-i\omega
  \left[\xi_{\rm eff}+\gamma_{\rm
      eff}\right]/\kappa}\label{eq:recursion3}.
\eee
where we left arguments implicit, and $\gamma_{\rm
  eff}=\gamma^{(+)}_n=\gamma^{(-)}_n$ (independent of $n$).  Thus,
\bee\label{eq:gammaeff}
\gamma_{\rm eff}=-\frac{\xi_{\rm eff}}{2}+ \Big{(}\frac{\kappa\xi_{\rm
    eff}}{-i\omega}\Big{)}^{1/2}\Big{(}1-\frac{i\omega \xi_{\rm
    eff}}{\kappa}\Big{)}^{1/2},
\eee
where we used the fact that the real part of $\gamma_{\rm eff}$ needs to be
positive (so the friction will dissipate energy) to choose the correct root of
the quadratic Eq. (\ref{eq:recursion3}). For small frequencies (long
times) we have 
\bee\label{eq:gammaeff_approx}
\gamma_{\rm
  eff}\sim \Big{(} \frac{\kappa \xi_{\rm
    eff}}{-i\omega}\Big{)}^{1/2}.
\eee
(B) From the self-consistency criterion we now determine $\xi_{\rm
  eff}$. Setting $n=0$ in Eqs. (\ref{eq:recursion1}) and
(\ref{eq:recursion2}) we obtain the generalized friction $\gamma_0^{(\pm
  )}=(\xi_{\rm eff}+\gamma_{\rm eff})/(1-i\omega \left[\xi_{\rm
    eff}+\gamma_{\rm eff}\right]/\kappa)$ for particle 0. Combining this
result with Eqs. (\ref{eq:mu_0}), (\ref{eq:EM_criterion}) and the
self-consistency criterion, $\bar{\mu}_0=\mu_{\rm eff}(\omega)$, we get:
\bee\label{eq:SC_crit1}
\int_0^\infty \frac{\prho(\xi_0)d\xi_0}{\xi_0+2\gamma_{\rm
    eff}}=\frac{1}{\xi_{\rm eff}+2\gamma_{\rm eff} }
\eee
where the right-hand side is the mobility obtained through the effective
medium equations [simply replace $\xi_0\rightarrow \xi_{\rm eff}$ and
$\gamma_0^{(\pm)}\rightarrow \gamma_{\rm eff}$ in Eq. (\ref{eq:mu_0})].  
Eq. (\ref{eq:SC_crit1})  can be rewritten:
\bee
\xi_{\rm eff} \int_0^\infty \frac{\prho(\xi_0)d\xi_0}{\xi_0+2\gamma_{\rm
    eff}}=\int_0^\infty \xi_0\frac{\prho(\xi_0) d\xi_0}{\xi_0+2\gamma_{\rm eff}}\label{eq:criterionexp}
\eee
where the normalization condition $1 = \int_0^\infty \prho(\xi_0)d\xi_0$ was
used. To summarize briefly, the effective friction $\xi_{\rm eff} (\omega)$ is
uniquely determined by Eqs. (\ref{eq:gammaeff})
[Eq. (\ref{eq:gammaeff_approx}) for small frequencies)] and
(\ref{eq:criterionexp}) for a given choice of $\prho(\xi_0)$.

Now we have to distinguish two cases: (\cone) the distribution $\prho(\xi_0)$
has a finite first moment, $\bar{\xi}=\int \xi_0 \prho(\xi_0) d\xi_0<\infty$. In
this case the $\xi_0$'s in the denominators on both sides of
Eq. (\ref{eq:criterionexp}) can be discarded for large $\omega$ since we
expect (and find self-consistently) that $\gamma_{\rm eff}(\omega)\to\infty$ as
$\omega\to 0$. For {\cone} systems we therefore have:
\bee\label{eq:C1_xi_eff_SI}
\xi_{\rm eff}(\omega)\sim \bar{\xi}.
\eee
which in the time-domain agrees with the result in the main text.
A more detailed analysis shows the result above applies for small frequencies
in the sense that $\omega\bar{\xi}/\kappa \ll 1$. 
(\ctwo) in the case of $\prho(\xi_0)$ not possessing a finite first moment we
assume a power-law tail $\prho(\xi_0)\sim A\xi_0^{-1-\alpha}$ for large $\xi_0$
($0<\alpha<1$). In this case the $\xi_0$ in the denominator on the left hand
side of Eq. (\ref{eq:criterionexp}) can again be neglected at small
$\omega$. In contrast, on the right hand side of Eq.
(\ref{eq:criterionexp}) the tail of $\prho(\xi_0)$ dominates the integral at
small $\omega$; this integral therefore needs to be evaluated. We then find,
using Eq. (\ref{eq:gammaeff_approx}), that Eq.
(\ref{eq:criterionexp}) becomes:
\bee
\xi_{\rm eff}\Big{(}\frac{4\kappa \xi_{\rm
eff}}{-i\omega}\Big{)}^{-1/2} = \frac{A\pi}{\sin(\pi \alpha)} \Big{(}\frac{4\kappa \xi_{\rm
eff}}{-i\omega}\Big{)}^{-\alpha/2},
\eee
so that for {\ctwo} systems the effective friction, for small frequencies,
decays according to a power-law:
\bee\label{eq:C2_xi_eff_SI}
\xi_{\rm eff}(\omega)\sim \frac{\chi}{(-i\omega)^{1-2\delta}}
\eee
with a prefactor $\chi=(4\kappa)^{1-2\delta}
(A\pi/\sin[\delta\pi/(1-\delta)])^{2(1-\delta)}$ and where
$\delta=\alpha/(1+\alpha)$. The result above
applies for the case that $\omega|\xi_{\rm eff}(\omega)|/\kappa \ll 1$.
 Fourier-transforming to the
time-domain we obtain Eq. (\ref{eq:C2_xi_eff}) in the main text.

Let us finally, for completeness, also give the expressions for the tracer
particle mobility $\mu_0(\omega)$. Combining Eqs. (\ref{eq:mu_0}) and
(\ref{eq:gammaeff_approx}) and taking the $\omega\rightarrow 0$ limit, we get
$\mu_0(\omega)\sim (-i\omega)^{1/2}/[2(\kappa\bar{\xi})^{1/2}]$ for {\cone}
probability densities, and
  \bee
\mu_0(\omega)\sim \frac{(-i\omega)^{1-\delta}}{2(\kappa\chi)^{1/2}}
  \eee
for {\ctwo} systems.

\section{Solution of the effective medium equations.}\label{eq:solution}

Taking the Fourier-transform with
respect to $n$ and $t$, the solution to Eq. (\ref{eq:EMH}) in the main text is
\bee\label{eq:y_q_w}
y(q,\omega)=\frac{\eta^{\rm eff}(q,\omega)+f(q,\omega)}{\kappa q^2-i \omega\xi_{\rm eff}(\omega)}
\eee
where $y_n(t)=x_n(t)-n\drho^{-1}$, and the Fourier-transforms of
with respect to $n$ of a function $B(n)$ is defined:
$B(q)=\int_{-\infty}^\infty e^{-iqn} B(n)dn$ with inverse
$B(n)=\int_{-\infty}^\infty e^{iqn} B(q)dq/(2\pi)$. Also Fourier-transforming
Eq. (\ref{eq:noise_corr}) of the main text leads to
\bea\label{eq:noise_corr_freq_space}
\langle \eta^{\rm eff}(q,\omega)\eta^{\rm eff}(q',\omega')\rangle&=& (2\pi)^2 k_B T[\xi_{\rm eff}(\omega)+\xi_{\rm eff}(\omega')]\nonumber\\
&&\times\delta(\omega+\omega')\delta(q+q'),
\eea
which together with
\bee\label{eq:white_noise}
\langle \eta^{\rm eff}(q,\omega)\rangle=0
\eee
provides a general solution to the effective medium harmonization equations.
Using Eq. (\ref{eq:y_q_w}) we find
\bea\label{eq:y_ny_n}
\langle y_n(t)y_{n'}(t')\rangle &=&\int_{-\infty}^\infty \frac{dq}{2\pi} \int_{-\infty}^\infty \frac{dq'}{2\pi} \int_{-\infty}^\infty\frac{d\omega}{2\pi} \int_{-\infty}^\infty \frac{d\omega'}{2\pi}\nonumber\\
&& \hspace{-2cm}\times  \frac{e^{i(qn+q'n')}e^{i(\omega t + \omega' t')}\langle\eta^{\rm eff}(q,\omega) \eta^{\rm eff} (q',\omega')\rangle}{[\kappa q^2-i\omega \xi_{\rm eff}(\omega)][\kappa q'^2-i\omega' \xi_{\rm eff}(\omega')]}
\eea
so that with the help of Eq. (\ref{eq:noise_corr_freq_space}) and some
simple algebraic manipulations we obtain
\bea\label{eq:y_ny_np}
\langle y_n(t)y_{n'}(t')\rangle &=&\frac{k_BT}{\kappa} \int_{-\infty}^\infty \frac{dq}{2\pi} e^{iq(n-n')} \int_{-\infty}^\infty\frac{d\omega}{2\pi} e^{-i\omega (t-t')} \nonumber\\
&&\times [\frac{1}{i\omega}\frac{1}{q^2-i\omega \xi_{\rm eff}(\omega)/\kappa}+\textrm{c.c.}]
\eea
where $\textrm{c.c.}$ denoted the complex conjugate. The relation above will
be used in subsequent sections.

\section{Dynamic structure factor}\label{sec:S_Q_t}

The dynamic structure factor is defined: $S(Q,t)=\sum_{n,m}\langle
e^{iQ[x_n(t)-x_m(0)]}\rangle /N$, where the sums extend over all particle
labels \cite{Berne_Pekora}. For an infinite, translationally invariant system
one can immediately perform one of the sums and get
\bee
S(Q,t)=\sum_{n=-\infty}^\infty \langle  e^{iQ[x_n(t)-x_0(0)]}\rangle .
\eee
Furthermore, assuming that the process is Gaussian one knows the
characteristic function and can therefore evaluate the average over noise to
find
\bee
S(Q,t)=\sum_{n=-\infty}^\infty e^{i Q \mu_n(t)-Q^2\sigma_n^2(t)/2}
\eee
where we have introduced the average and variance
\begin{eqnarray}
\mu_n(t)&=&\langle x_n(t)-x_0(0) \rangle=n\drho^{-1} ,\\
\sigma_n^2(t)&=&\langle  [x_n(t)-x_0(0)-\mu_n(t)]^2\rangle \nonumber\\
&=&\langle  [y_n(t)-y_0(0)]^2\rangle.
\end{eqnarray}
The variance $\sigma_n^2(t)$ can be evaluated using Eq.
(\ref{eq:y_ny_np}) as
\begin{eqnarray}
\sigma_n^2(t)&=&\int\frac{d q}{2\pi}\int\frac{d\omega}{2\pi} \left(1-e^{i q n-i\omega t}\right)\nonumber\\
&&\times\left(\frac{k_B T}{\kappa q^2}\frac{1}{\kappa q^2/\xi_{\rm eff}(\omega)-i\omega}+\textrm{c.c.}\right)
\end{eqnarray}
At this point we introduce a collective diffusion constant $\lambda_c$
by writing $\xi_{\rm eff}(\omega)=\kappa \lambda_c^{-1} (-i
\omega)^{2\delta-1}/\drho^2$. For {\cone} systems we see from Eq.
(\ref{eq:C1_xi_eff_SI}) that $\delta=1/2$ and
$\lambda_c=\kappa/(\drho^2 {\bar \xi})$, and for {\ctwo} systems Eq.
(\ref{eq:C2_xi_eff_SI}) gives
$\lambda_c=\kappa/(\rho^2\chi)$. Now we perform the
$\omega$-integration by changing variable $-i\omega\to s$ and using that a
definition of the Mittag-Leffler function $E_\alpha(\cdot)$ is
\bee
\int_0^\infty E_\alpha(-\theta t^\alpha)e^{-s t} d t=\frac{s^{\alpha-1}}{s^\alpha+\theta},
\eee
to find
\bee
\sigma_n^2(t)=2\int\frac{d q}{2\pi}\frac{k_B T}{\kappa q^2}(1-e^{i q n}E_{2\delta}(-\drho^2 \lambda_c q^2 |t|^{2\delta})).
\eee
This result can also be written
\bee
\sigma_n^2(t)=\frac{k_B T}{\kappa}\left(|n|+{\drho \sqrt{\lambda_c}|t|^\delta} h(n/(\drho\sqrt{\lambda_c}|t|^\delta))\right) ,
\eee
where we have introduced the function
\bee
h(z)=2\int\frac{d q}{2\pi}\cos(q z)[1-E_{2\delta}(-q^2)]/q^2 .
\eee
Note that $h(z)$ is bounded from above by $h(0)$.

To get further, we are now going to take the hydrodynamic limit of $Q\to 0$
and $t\to\infty$. In taking this limit we keep $Q t^\beta$ fixed, where
$\beta$ is an exponent which will be chosen such that we get a non-trivial
result. If we make the ansatz that $\beta>\delta/2$ then $Q^2 t^\delta\to 0$
and since $h(z)$ was bounded from above we have then to first order in $Q^2
t^\delta$
\bea
e^{-\frac{1}{2}Q^2\sigma_n^2(t)}\sim e^{-\frac{1}{2}Q^2\frac{k_B T}{\kappa}|n|}&&\\
\times\left(1-\frac{1}{2}Q^2 \frac{k_B T}{\kappa}\drho\sqrt{\lambda_c}|t|^\delta h(n/(\drho \sqrt{\lambda_c} |t|^\delta) \right) .
\eea
If we insert this into the expression for $S(Q,t)$ and take the continuum
limit by replacing the sum $\sum_n$ with an integral $\int d n$ then the
expression for $S(Q,t)$ can now be evaluated. For $\beta<\delta$ one gets
$S(Q,t)\sim 0$ while for $\beta>\delta$: $S(Q,t)\sim k_B T\drho^2/\kappa$. The
non-trivial result appears for $\beta=\delta$ where one finds
\bee
S(Q,t)\sim \frac{k_B T \drho^2}{\kappa}E_{2\delta}(-\lambda_c Q^2 |t|^{2\delta}) .\label{eq:SSQt}
\eee
This is the expression for $S(Q,t)$ used in the main text.

As mentioned in the text, the Mittag-Leffler decay of the structure
factor implies, by Onsager's regression hypothesis \cite{onsager31}, that
perturbations of the concentration $c(X,t)=\langle \sum_n \delta(X-x_n(t)
\rangle$ decay according to the fractional diffusion equation
(FDE). To see this note that the structure factor is the Fourier transform of
density-density correlations. Onsager's regression hypothesis (alternatively
the fluctuation-dissipation theorem) implies that perturbations in the density
decay in the same way as correlations. Thus we have for the Fourier transform
$c(Q,t)=\int d X\,e^{-i Q X}c(X,t)$ that it decays according to
$c(Q,t)=c(Q,t=0)\times E_{2\delta}(-\lambda_c Q^2 |t|^{2\delta})$ ($Q>0$). This
implies that it obeys a fractional relaxation equation
\bee
\frac{\partial}{\partial t}c(Q,t)=-\lambda_c Q^2 {}_0D_t^{1-2\delta} c(Q,t) ,\label{eq:FDEQ}
\eee
where the fractional Riemann-Liouville operator ${}_0D_t^{\alpha}$ is
defined for $0<\alpha<1$ by
\bee
{}_0D_t^\alpha c(Q,t)=\frac{1}{\Gamma(1-\alpha)}\frac{\partial}{\partial t}\int_0^t dt' \frac{c(Q,t')}{(t-t')^\alpha}.
\eee
%
%
Taking the inverse Fourier transform of Eq. (\ref{eq:FDEQ}) one arrives
at the FDE
\bee
\frac{\partial}{\partial t}c(X,t)=\lambda_c\, {}_0D_t^{1-2\delta} \frac{\partial^2}{\partial X^2}c(X,t).\label{eq:FDE}
\eee
given in the main text.

\section{Tracer particle mean square displacement.}\label{sec:tracer_MSD} 
Let us now derive an expression for the MSD in the absence of an external force. We have
\bea
\langle \delta x_\mathcal{T}(t)^2 \rangle &=&
\langle [y_0(t)-y_0(0)]^2\rangle\nonumber\\
&=&2[\langle y_0(0)^2\rangle -\langle
y_0(t)y_0(0)\rangle ]
\eea
due to stationarity. Here $\mathcal{T}=0$ has been chosen without lack of
generality.  Using
Eq. (\ref{eq:y_ny_np}) and the integral
$\int_{-\infty}^\infty (q^2+C)^{-1}dq = \pi/\sqrt{C}$ for ${\rm Re}[C]>0$, we
can express the time-derivative of the MSD according to:
\bea
\frac{d}{dt}\langle \delta x_\mathcal{T}^2 \rangle &=&\frac{k_BT}{2\sqrt{\kappa}}\int_{-\infty}^\infty d\omega e^{-i\omega t} \nonumber\\
&&\hspace{-1cm}\times \Big{(} \frac{1}{[-i\omega \xi_{\rm eff} (\omega)]^{1/2}}-\textrm{c.c.}\Big{)}
\eea
Finally, rewriting the expression above as a sine-transform, inserting
Eqs. (\ref{eq:C1_xi_eff_SI}) and (\ref{eq:C2_xi_eff_SI})
 and integrating [using the fact that $\langle \delta x_\mathcal{T} (0)^2
\rangle=0$] we recover the MSDs given in the main text. 

\begin{figure}[tb]
\begin{center}
\includegraphics[width=8.5cm]{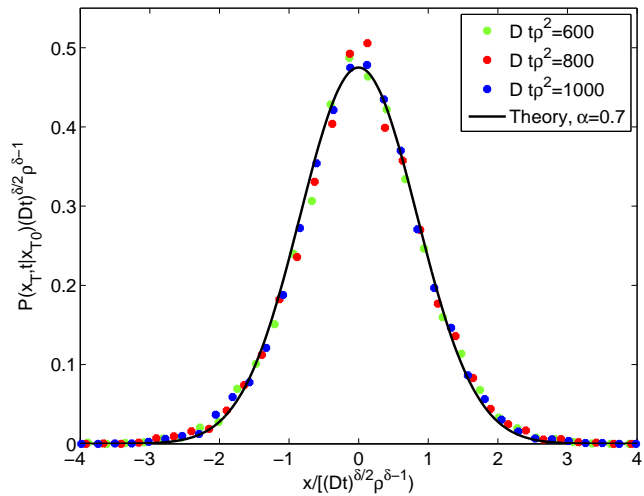}
\end{center}
\caption{Collapse-plot for the tagged particle probability density
    function at different times. The solid black curve is a zero-mean
    Gaussian with a width given by Eq. (\ref{eq:MSD}). The symbols correspond
    to heterogeneity-averaged simulations over 8000 noise realizations with
    $N=501$ particles in a system of length L=5001b ($b$ is the particle size)
    with $\alpha=0.7$.
  \label{fig:pdf}}
\end{figure}

 The noise is assumed to be Gaussian distributed, and therefore the tracer
  particle PDF is a Gaussian with zero mean and a width determined by the MSDs
  given in Sec. \ref{sec:tracer}; Fig. \ref{fig:pdf} compares
  heterogeneity-averaged simulation results for the PDF to this prediction,
  finding excellent agreement.

\section{Tracer particle dynamics in the presence of a time-varying external
  force.}\label{sec:force}

To test the effective medium harmonization approach further we have considered
the response of a tagged particle to an external force.  For this purpose let
us assume that a harmonically oscillating force
$f_n(t)=\delta_{n,0}F_0\cos(\omega_0 t)$ acts on particle $0$, where
$\omega_0$ is the angular velocity and $F_0$ the force amplitude.  From the
formal solution Eq. (\ref{eq:y_q_w}) and Eq. (\ref{eq:white_noise})
we find:
\bee\label{eq:y_q_w_f}
y(q,\omega)=\frac{\pi F_0[\delta(\omega-\omega_0)+\delta(\omega+\omega_0)]}{\kappa q^2-i \omega\xi_{\rm eff}(\omega)}.
\eee
Further, by performing the inverse Fourier-transform with respect to $q$, and
setting $n=0$, we get
\bee\label{eq:y_0}
\langle y_0(t)\rangle_f =\frac{F_0}{2\kappa} {\rm
  Re} \left( \frac{e^{i\omega_0t}}{[-i\omega_0\xi_{\rm eff}(\omega_0)/\kappa]^{1/2}}\right) ,
\eee
where ${\rm Re}\{..\}$ denotes the real part and the subscript indicates an
average in the presence of a force. For a {\cone} [{\ctwo}] system we use the
friction constant in Eq. (\ref{eq:C1_xi_eff_SI}) [Eq.
(\ref{eq:C2_xi_eff_SI})] in the expression above.  Switching to the real
tagged particle coordinate $x_\mathcal{T}$ assuming $\langle
x_\mathcal{T}(t=0)\rangle=0$ we get $\langle
x_\mathcal{T}(t)\rangle_f=[F_0/(2\sqrt{\kappa \bar{\xi} \omega_0})]
\cos(\omega_0 t-\phi)$ where $\phi=\pi/4$ for {\cone} systems. For {\ctwo}
systems we find
\bee \langle
\delta x_\mathcal{T}(t)\rangle_f=\frac{F_0}{2\sqrt{\kappa \chi}
  \omega_0^{\delta}}\cos(\omega_0 t-\phi)
\label{eq:force_resp_app}
\eee
with the non-trivial phase-shift $\phi=\delta
\pi/2<\pi/4$.

\section{Generalized Einstein relation.}\label{sec:gen_Einstein}

Suppose a static force $F_0$ is applied to particle $0$ at time $t=0$,
i.e. $f_n(t)=F_0\delta_{n,0}\theta(t)$.  From the fluctuation-dissipation
theorem \cite{kubo66}, we have that the following simple relation holds
between the tagged particle mean square displacement in the absence of the
force $\langle \delta x_\mathcal{T}^2 (t) \rangle$, and the mean displacement
in the presence of the force $\langle \delta x_\mathcal{T} (t)\rangle_F$
\bee\label{eq:gen_Einstein}
\langle \delta x_\mathcal{T} (t)\rangle_f = \frac{F_0}{2 k_B T} \langle
\delta x_\mathcal{T}^2 (t) \rangle
\eee
which is referred to as the generalized Einstein relation. A consistency check
of the effective medium-harmonization formalism is thus to check that this
relation holds. From Eq. (\ref{eq:y_q_w}) we straightforwardly compute
the average velocity (in frequency space): $\langle v_n(\omega)\rangle
=-i\omega \langle y_n(\omega)\rangle$ for the type of force considered
here. Further, by Fourier-inverting and setting $n=\mathcal{T}=0$, and using
Eq. (\ref{eq:C2_xi_eff_SI}) we find
\bee
\langle v_\mathcal{T}(t)\rangle = \frac{F_0}{2\sqrt{\kappa\chi}} \frac{t^{\delta-1}}{\Gamma(\delta)}\theta(t)
\eee
for {\ctwo} systems. By integrating the expression above over time 
and comparing with the MSD in the main text we
find that Eq. (\ref{eq:gen_Einstein}) is satisfied for $t>0$. A similar
calculation for {\cone} systems shows that indeed the generalized Einstein
relation is satisfied also then.

\section{Simulation scheme}\label{sec:simulation_details} 

We consider $N$ hardcore particles of linear size $b$ in a box of length $L$
(the box extends from $-L/2$ to $L/2$). At random times the particles performs
hops with the jump length taken from a Gaussian distribution of width
$a$. Particle $n$ ($n=1,...,N$) has a hop rate $q_n$, where $q_n$ is related
to the friction constant $\xi_n$ by $q_n=2 k_B T/(\xi_n a^2)$. We are
interested in the stochastic dynamics of the center positions $X_n$ of all
particles, when we have the single-file constraint $X_{n+1}-X_n>b$ at all
times. Also, particles cannot move out of the box: $X_1>-(L-b)/2$ and
$X_N<(L-b)/2$. Our numerical approach detailed below,
builds on a previous scheme for lattice simulations called the trial-and-error
Gillespie algorithm \cite{TA_etal}. Our continuum version of that algorithm
is:

\begin{enumerate}

\item Generate the cumulative sums of the free total rate constants
\bea\label{eq:p_mu}
p_0&=&0,\nonumber\\
p_n&=&\sum_{m=1}^n q_m, \qquad n=1,...,N.
\eea
\item Generate a random initial configuration of particle positions $X_n(0)$
($n=1,...,N$). This step requires some care, see next
section.

\item Set the time equal to zero, $t=0$.

\item Draw a uniform random number $r$ ($0<r<1$). Construct a waiting time $\tau$ according to:
\bee
\tau= \frac{1}{p_N} \log \left( \frac{1}{r} \right),
\eee 
i.e. $\tau$ is taken from an exponential distribution
$\Phi(\tau)\propto\exp(-\tau \sum_{n=1}^N q_n)$. Update the time $t\rightarrow
t+\tau$.

\item Draw a new uniform random number $r$ ($0<r<1$). A trial particle $n$ is
  determined by the $n$ which satisfies
\bee\label{eq:ineq}
p_{n-1} \le r p_N <  p_n
\eee
i.e., particle $n$ is chosen with a probability $\propto q_n$.

\item Draw a random attempt jump length $l$ from the Gaussian distribution of
  width $a$: $P(l)=(2\pi a^2)^{-1/2}\exp[-(l-\mu)^2/(2a^2)]$. For the case of
  a harmonically oscillating external force on the tracer particle
  $f_n(t)=\delta_{n,0} F_0\cos(\omega_0 t)$,
  $\mu=\alpha_{0}\cos[\omega_{0}t])$ for the tracer particle with
  $\alpha_0=F_0/(\xi_0 q_0)$, otherwise $\mu=0$.

\item Loop over the neighboring particles in the direction of the attempted
jump, and find the number $M$ of particles which have their center
positions within a distance $l-b$ of their previous neighbor.
 
\item Draw a uniform random number $r$ and move particle $n$ and the $M$
  neighbor particles a distance $l$ if $r<\xi_n/\sum_{m=n}^{n\pm M}\xi_m$
  (the sum ranges from $n$ to $n+M$ if the attempted jump is to the right,
  otherwise the range is $n-M$ to $n$), and there is no wall preventing the
  move. If the move occurs then update the particle positions.

\item Return to step (4).

\end{enumerate}
The scheme above produces a stochastic time series $X_n(t)$, and if
steps (2)-(9) are repeated many times, ensemble averages
can be computed. Note that step (1) does not have to be repeated for
each simulation run (each realization of the noise) for the {\em
  non-averaged} case, i.e. if the free total hop rates are the same for
all simulation runs. For the {\em heterogeneity-averaged} case, where we assign new
hop rates for the particles for every new simulation run, step 1 has
to be repeated for each one of them. An efficient method for
performing the search for the $n$ satisfying the inequality in
Eq. (\ref{eq:ineq}) was presented previously \cite{TA_etal}. We here
point out that in our previous study \cite{TA_etal} we instead of
Eq. (\ref{eq:ineq}) used $p_{n-1} < r p_N \le p_n$. This criterion for
choosing a trial particle causes the algorithm to get stuck for the
(exceptional) case that $p_{n-1}\equiv rp_N$ if combined with the
search algorithm in appendix B of Ref. \onlinecite{TA_etal}. We are grateful
to Karl Fogelmark for pointing this out to us.

In step 2 in our algorithm, we set the initial positions randomly (with the
possible additional constraint of a fixed position for a tracer particle). For
point-particles this present no problem - for completely random initial
positioning one simply needs to draw $N$ uniform random
numbers between $-L/2$ and $L/2$, which are then sorted and assigned as
initial positions. For finite size particles one must make sure that there is
no overlap of the particles. The same procedure as used above, with the
additional step to discard configurations where two or more particle overlap,
is very inefficient for large $N$. We instead use the following mapping
between the finite-size particle problem [initial positions $X_n(0)$,
$n=1,...,N$] and a point-particle problem [initial positions $\tilde{X}_n(0)$]
\cite{LIAM1,LIAM2} 
\bea
L&=&l+Nb,\nonumber\\
X_n(0)&=&\tilde{X}_n(0)+nb-\frac{N+1}{2}b.  
\eea 
Step 2 in the simulation scheme then becomes: draw random numbers between
$-l/2$ and $l/2$ which, after sorting, gives $\tilde{X}_n(0)$. Then use the
relation above to determine $X_n(0)$. This procedure is used in the
simulations providing the dynamic structure factor. For the tracer particle
simulations we fix the initial position of a tracer particle at the center of
the system with equally many particles to the left and right. The procedure
above then has to be done separately for the particles to the left and right
of the tracer particle.

Step 7 and 8 constitute our rule for handling collisions between particles.
These rules are based on the idea of having momentum conservation on average
in collisions while at the same time fulfilling detailed balance. The total
momentum dissipated to the surrounding medium if particle $n$ moves a distance
$l$ is $l\xi_n$. When the $M+1$ particles move the dissipated momentum is
$l\sum_{m=n}^{n\pm M}\xi_m$. Carrying out the later move with a probability
$\xi_n/\sum_{m=n}^{n\pm M}\xi_m$ makes the dissipated momentum independent, on
average, of whether collisions happen or not. Detailed balance is satisfied,
because the rate for the attempted move of the $M+1$ particles is $q_n\propto
1/\xi_n$. Thus the rate at which the particles actually move will be
proportional to $1/\xi_n\times\xi_n/\sum_{m=n}^{n\pm
  M}\xi_m=1/\sum_{m=n}^{n\pm M}\xi_m$, and identical to the rate with which
the reverse move (particle $n\pm M$ initiating the reverse move of the $M+1$
particles in the opposite direction) occurs, since in equilibrium all allowed
positions of the hard-core interacting particles are equally
probable. Detailed balance is therefore satisfied by the above algorithm.

In the simulations of {\ctwo} systems we draw the friction constants $\xi_n$
from the density [$0<\alpha<1$]
\bee\label{eq:rho}
\prho(\xi_n)=A \xi_n^{-1-\alpha}\
{\rm for}\ \xi\ge \xi_c
\eee
and $\prho(\xi_n)=0$ for $\xi_n<\xi_c$. The normalization constant is
$A=\alpha\xi_c^\alpha$. In order to generate a random friction constant
$\xi_n$ according to the density above, in a standard fashion
\cite{gillespie76}, we set the cumulative distribution
$C(\xi_n)=\int_0^{\xi_n} \prho(\xi')d\xi'$ equal to a uniform random number
$r$ between 0 and 1, $C(\xi_n)=r$. Explicitly we then get
\bee\label{eq:xi}
\xi_n=\xi_c (1-r)^{-1/\alpha}.  
\eee 
In each simulation we determine a random friction constant $\xi_n$ for
each particle using Eq. (\ref{eq:xi}). Then, using the fact that the
free particle diffusion constant $D_n$ is related to the friction and
hop rates as $D_n=q_na^2/2=k_BT/\xi_n$, we get random hop rates:
$q_n=2k_BT/(\xi_n a^2)$ used in the simulations. We fix the
normalization constant $A$ by setting the average diffusion constant
$\bar{D}=k_BT \int \xi_n^{-1}\prho(\xi_n)d\xi_n=\alpha k_B
T/[(1+\alpha)\xi_c]$ to unity in the simulations.

In all simulations except those for the structure factor $S(Q,t)$ we have used
a width $a=1$ for the jump length distribution. For the structure factor
simulations $a=0.25$ was used due to a higher choice of density
$\drho$. Ideally $a$ should be as small as possible compared with the average
distance between particles $\drho^{-1}$ to approximate the diffusive limit of
 moving particles in a spatial continuum.

%
%


\begin{thebibliography}{10}

\bibitem{Pearson}
K.~Pearson.
\newblock {\em Nature}, 72(1865):294, 1905.

\bibitem{Lord_Rayleigh}
L.~Rayleigh.
\newblock {\em Nature}, 72(1866):318, 1905.

\bibitem{Fisher_84}
M.E. Fisher.
\newblock {\em Journal of Statistical Physics}, 34(5):667--729, 1984.

\bibitem{KKDGPRSUK}
V.~Kukla, J.~Kornatowski, D.~Demuth, I.~Girnus, H.~Pfeifer, L.V.C. Rees,
  S.~Schunk, K.K. Unger, and J.~K\"arger.
\newblock {\em Science}, 272(5262):702, 1996.

\bibitem{meersmann00}
T.~Meersmann, J.W. Logan, R.~Simonutti, S.~Caldarelli, A.~Comotti, P.~Sozzani,
  L.G. Kaiser, and A.~Pines.
\newblock {\em J. Phys. Chem. A}, 104(50):11665--11670, 2000.

\bibitem{hahn96}
K.~Hahn, J.~K{\"a}rger, and V.~Kukla.
\newblock {\em Physical Review Letters}, 76(15):2762--2765, 1996.

\bibitem{WBL}
Q.H. Wei, C.~Bechinger, and P.~Leiderer.
\newblock {\em Science}, 287(5453):625, 2000.

\bibitem{gupta95}
V.~Gupta, S.S. Nivarthi, A.V. McCormick, and H.~Ted~Davis.
\newblock {\em Chemical Physics Letters}, 247(4-6):596--600, 1995.

\bibitem{HOKE}
AL~Hodgkin and RD~Keynes.
\newblock {\em The Journal of Physiology}, 128(1):28, 1955.

\bibitem{Elf_09}
G.W. Li, O.G. Berg, and J.~Elf.
\newblock {\em Nature Physics}, 5(4):294--297, 2009.

\bibitem{LE}
D.G. Levitt.
\newblock {\em Physical Review A}, 8(6):3050--3054, 1973.

\bibitem{Arratia_83}
R.~Arratia.
\newblock {\em The Annals of Probability}, 11(2):362--373, 1983.

\bibitem{HA}
T.E. Harris.
\newblock {\em Journal of Applied Probability}, 2(2):323--338, 1965.

\bibitem{Alexander_Pincus_78}
S.~Alexander and P.~Pincus.
\newblock {\em Physical Review B}, 18(4):2011--2012, 1978.

\bibitem{Kollmann_03}
M.~Kollmann.
\newblock {\em Physical Review Letters}, 90(18):180602, 2003.

\bibitem{barkai2009theory}
E.~Barkai and R.~Silbey.
\newblock {\em Physical Review Letters}, 102(5):50602, 2009.

\bibitem{LIAM2}
L.~Lizana and T.~Ambj{\"o}rnsson.
\newblock {\em Physical Review E}, 80:051103, 2008.

\bibitem{lizana_10}
L.~Lizana, T.~Ambj\"ornsson, A.~Taloni, E.~Barkai, and M.A. Lomholt.
\newblock {\em Physical Review E}, 81:051118, 2010.


\bibitem{Khokhlov}
 A.Y. Grosberg and A.R. Khokhlov. {\em Statistical Physics of Macromolecules},
 AIP Press, New York, 1994.

\bibitem{landauer52}
R.~Landauer.
\newblock {\em Journal of Applied Physics}, 23:779, 1952.

\bibitem{kirkpatrick71}
S.~Kirkpatrick.
\newblock {\em Physical Review Letters}, 27(25):1722--1725, 1971.

\bibitem{alexander81}
S.~Alexander, J.~Bernasconi, WR~Schneider, and R.~Orbach.
\newblock {\em Reviews of Modern Physics}, 53(2):175--198, 1981.

\bibitem{Jara_Gonzalves}
P.~Gon{\c{c}}alves and M.~Jara.
\newblock {\em Journal of Statistical Physics}, 132(6):1135--1143, 2008.

\bibitem{Flomenbom}
O. Flomenbom, Phys. Rev. E {\em 82}, 031126 (2010).

\bibitem{TA_etal}
T.~Ambj{\"o}rnsson, L.~Lizana, M.A. Lomholt, and R.J. Silbey.
\newblock {\em The Journal of Chemical Physics}, 129:185106, 2008.

\bibitem{Jara_09}
M.~Jara.
\newblock {\em E-print: arXiv:0901.0229}, 2009.

\bibitem{kubo66}
R.~Kubo.
\newblock {\em Rep. Prog. Phys.}, 29:255, 1966.

\bibitem{Berne_Pekora}
B.J. Berne and R.~Pecora. {\em Dynamic Light Scattering with applications to
chemistry, biology and physics}, Dover Pubns, 2000.

\bibitem{MEKL1}
R.~Metzler and J.~Klafter.
\newblock {\em Physics Reports}, 339(1):1--77, 2000.

\bibitem{onsager31}
L.~Onsager.
\newblock {\em Phys. Rev.}, 37:405, 1931.

\bibitem{weiss_09}
J.~Szymanski and M.~Weiss.
\newblock {\em Physical Review Letters}, 103(3):38102, 2009.

\bibitem{kirkpatrick73}
S.~Kirkpatrick.
\newblock {\em Percolation and Conduction}, 45:574--588, 1973.

\bibitem{LIAM1}
L.~Lizana and T.~Ambj{\"o}rnsson.
\newblock {\em Physical Review Letters}, 100(20):200601, 2008.

\bibitem{gillespie76}
D.T. Gillespie.
\newblock {\em J. Comp. Phys.}, 22:403, 1976.

\end{thebibliography}
\end{document}